\documentclass[USenglish]{article}

\usepackage[utf8]{inputenc}				%(only for the pdftex engine)
\usepackage[big,online]{dgruyter}	%small,big | online,print,work
\usepackage{lmodern}
\usepackage{microtype}
\usepackage[numbers,square,sort&compress]{natbib}
\usepackage[cmyk]{xcolor}
\definecolor{blue}{cmyk}{1,0,0,0}

\theoremstyle{dgthm}

\theoremstyle{dgdef}

\def\be{\begin{equation}}   \def\ee{\end{equation}}   \def\bea{\begin{eqnarray}}    \def\eea{\end{eqnarray}}  
    \def\d{{\rm d}}

\def\l{\left}
\def\r{\right}

\def\p{\partial}

\begin{document}

%%%--------------------------------------------%%%
	\articletype{Research Article}
	\received{Month	DD, xxxx}
	\revised{Month	DD, xxxx}
  \accepted{Month	DD, xxxx}
  \journalname{J.~Non-Equilib.~Thermodyn.}
  \journalyear{xxxx}
  \journalvolume{xx}
  \journalissue{X}
  \startpage{1}
  \aop
  \DOI{10.1515/jnet-xxxx-XXXX}
%%%--------------------------------------------%%%

\title{Effect Structure and Thermodynamics Formulation of Demand-side Economics}

\runningtitle{Effect Structure and Thermodynamics Formulation of Demand-side Economics}
%\subtitle{Insert subtitle if needed}

\author[1]{Burin Gumjudpai}
\runningauthor{B. Gumjudpai}
\affil[1]{\protect\raggedright
Centre for Theoretical Physics \& Natural Philosophy ``Nakhonsawan Studiorum for Advanced Studies", Mahidol University, Nakhonsawan Campus,  Phayuha Khiri, Nakhonsawan 60130, Thailand, e-mail: burin.gum@mahidol.ac.th}

%\communicated{...}
%\dedication{...}
	
\abstract{We propose concept of equation of state (EoS) effect structure in form of diagrams and rules. This concept helps justifying EoS status of an empirical relation. We apply the concept to closed system of consumers and we are able to formulate its EoS.  According to the new concept, EoS are classified into three classes.  Manifold space of thermodynamics formulation of demand-side economics is identified.  Formal analogies of thermodynamics and economics consumers' system are made. New quantities such as total wealth, generalized utility and generalized consumer surplus are defined. Microeconomics' concept of consumer surplus is criticized and replaced with generalized consumer surplus. Smith's law of demand is included in our new paradigm as a specific case resembling isothermal process.
Absolute zero temperature state resembles the nirvana state in Buddhism philosophy.  Econometric modelling of consumers' EoS is proposed at last.}


\maketitle

\date{\today}

\section{Introduction} \label{Intro}

For a system of many collective agents, postulating on existence of equilibrium state is the first step for further theoretical building. Thermodynamics describes such systems averagely and aggregately. Microscopic description is linked by statistical mechanics. Statistical properties of simple microscopic system depends on its properties, e.g. mass, charge and spin and on how these agents interact to each other.  Distinct microscopic properties correspond to distinct statistical properties  represented with distribution functions. In social systems, we do not have knowledge that is analogy to mechanics, that is the equation of motion. Social agents are choice-making entities. Their activities of our interests are not about motion, but about how they behave and how they make choices.
Despite lacking of analogous concept of mechanics, social systems are collective and aggregated as of  gas molecules.
In thermodynamics, it is possible to control experiments and test them hence rather precise empirical laws can be formulated.
On the other hand, social systems are open and easily disturbed. Therefore it is hard to find a precise law. It is difficult for ones to  profoundly scrutinize data and find its true fundamental principles. Assumptions are made so that phenomena could be described under some certain conditions.  Although, social sciences, e.g. economics seem to be highly phenomenological, the fact that they are collective and aggregated like thermodynamics suggests that there should be some common fundamental descriptions. Ones might argue that idealized concepts could not predict real-world open economy, however  this does not imply that their common theoretical background does not exist. In physics, ideal system is usually studied before extending  to real-world descriptions with add-on terms or coefficients. We shall apply this procedure of knowledge building in thermodynamics to system considered in microeconomics. We consider consumption-side economy here.  The term ``consumption'' is prefer to ``demand'' because consumption has saturated meaning. However following economics convention, both are used.
We revise and give extensions to results in \cite{BGDemand, BGThesis, BG3, BG2}. Earlier attempt was reported in \cite{BG1} of which inconsistent results found therein illuminate way and insight to our proceeding theoretical building. We speculate that thermodynamics should be fulfilled with additional concept of {\it effect structure} which proposes new characters of any equations of state (EoS). This helps us describing EoS more precisely. It helps classifying EoS and justifying empirical formula for a status of an EoS.  Realizing that  thermodynamics is natural, its paradigms are hold as foundation in looking at economics system. We deploy the new concept of effect structure in justifying of thermodynamics-analogous manifold space variables for demand-side economy. This provides space for its EoS.
Some common characteristics of thermodynamics and microeconomics are such as
\begin{enumerate}
  \item collective and aggregated nature of system,
  \item existence of equilibrium states determined with dual state variables,
  \item existence of conservative constraint,
  \item in approaching equilibrium, there is some preference for one quantity not to decrease and another not to increase.
\end{enumerate}
We suggest further that other aspects of  thermodynamics should be possessed by microeconomics. These are
\begin{enumerate}
  \item existence of infinite thermal and entropic states,
  \item existence of EoS, $g(X_i, Y_i, T) = 0$,
  \item a scalar derived from conservative quantity (internal energy, $U$) playing role of temperature,
  \item $U$ is minimized and $S$ is maximized at equilibrium,
   \item absolute values of $U$ and of entropy $S$ in real world systems are unmeasurable (only relative values are measurable),
   \item existence of truly endogenous function and effect structure in the EoS.
 \end{enumerate}
In building thermodynamics formulation of economics,  we impose these six additional aspects of thermodynamics into our theoretical building.
Existence of EoS surface is equivalent to existence of equilibrium states and $2(n+1)$ tuples manifold coordinate space $\{X_i, Y_i, S, T\}$. Extensive and intensive coordinates $X_i, Y_i$  (where $i = 1 \ldots n$) are {\it mechanical pair variables} - dual states for the work term. On the EoS surface (known as equilibrium surface), $S$ is fixed, hence left with $(2n+1)$ degrees of freedom. Equilibrium thermodynamics space is therefore $\mathcal{M} = \{X_i, Y_i, T\}$. Existence of infinite thermal and entropic states is imposed by Carathéodory’s empirical axioms (see, e.g. M\"{u}nster 1970  \cite{munster}, Land\'{e}  1926  \cite{cara}, Buchdahl 1955 \cite{cara2},  Turner  1960 \cite{cara3}
and Buchdahl 1966 \cite{cara4}).
Existence of thermal state $t(X_i, Y_i)$ alone implies the EoS, $g(X_i, Y_i, T) = 0$ to exist, for instance, $g(V,P,T)=0$. {\it Field} functions, such as $U$, are dubbed thermodynamics potentials which are defined on the EoS surface. Different potentials are related via Legendre transformation based on different thermodynamics sets of coordinates.  Energy transfer are the work, $\delta W_i = Y_i \d X_i$ and the heat, $\delta \mathcal{Q} = T \d S$ terms. Both are inexact differentials. Considering {\it stock} and {\it flow} types economics variables, it is obvious that flow variables resemble energy transfer, whereas stock variables are state variables, yet to be classified as either potentials or manifold coordinates.  This lack of completion is because, in economics, there is neither concept of EoS nor potential - at least in the same spirit of thermodynamics (see, e.g. Debrue 1972 \cite{debrue} and Mas-Colell 1985 \cite{Mas}).
 As mentioned, manifold space for an thermodynamics EoS is  $\{X_i, Y_i, T\}$.  Manifold space of consumption-side economics for performing constrained optimization is the set of consumption quantities $\{Q^{\d}_i\}$, of type $i$ commodity. In 2008, Smith (a physicist) and Foley (an economist) \cite{Smit} proposed consumers' EoS in form of $g(Q^{\d}, Pr, \rm{MRS}) = 0$ where $Q^{\d}, Pr, \rm{MRS}$ are consumption quantity, price and marginal rate of substitution. Considering multiple types of commodity, demand functions $Q^{\d}_i$ are optimized with consumers' budget conservative constraint so that the equilibrium price is settled and the utility function, $\mathcal{U}$ is obtained. This rather differs from classical thermodynamics of which optimization is performed on $U$ and $S$ functions by the second law while the conservative constraint is the first law.

Investigation of thermodynamics and economics connection could date back to 1892  when Fisher's thesis
considered $\p \mathcal{U}/\p Q^{\d}_i $ as analogy to $\p U/\p X_i$, $\p U/\p Y_i$, $\p U/\p T$ and $\p U/\p S $ \cite{Fish}.
In 1909  Walras considered mechanical gradient of potential energy in analogy to $\p \mathcal{U}/\p Q^{\d}_i $  \cite{walras} .
Other investigations are such as Lisman's 1949 \cite{Lisman}) and Saslow 1999 \cite{Sasl}, however, in both works, EoS are not considered and their concepts are different from ours.  In 1960, this stream of thought was realized wider by Samuelson’s critique of why lacking of identification of economics quantities in the realm of thermodynamics  could {\it ``lead anyone to overlook or deny the mathematical isomorphism that does exist between minimum systems that arise in different disciplines''} \cite{Samu}. In 2006, Sousa and Domingos considered formal approach to economics and they showed that generalized Le Chatelier principle is indeed economics integrability hence resulting in existing of intensive variable ($\pi_i$) as function of multiple extensive variables, $\pi_i = \pi_i(X_1,\ldots X_i \ldots X_n) $  \cite{sousa2006}. In  their work, this relation is regarded as an equation of state, however the EoS in this form does not resemble the spirit of thermodynamics because it does not have the concept of empirical dual states of mechanical pairs $(X_i, Y_i)$ and thermal-entropic pair $(S, T)$, such that the work and heat terms can be expressed. Thermodynamics EoS should include thermal state, $T$ in  the argument of EoS function. In such framework,  $(T,S)$ are treated  on equal basis of normal empirical intensive and empirical extensive variables
rather than realizing that $T$ and $S$ are postulated variables given by Carathéodory’s empirical axioms. In 2010, Arroyo Col\'{o}n viewed economy as system of heat engines.  Consumption  and production  are considered as transferring of commodity quantities into and out-of some opposite sides. However, the concept of EoS is not mentioned either \cite{Colon}.  Here, we realize that consumption process is physically distinct from production process just as how hydrostatics system is different from paramagnetic substance. Producers and consumers are agents that behave solely with different roles, although they might be the same group of people.
Our work here is a formal approach, not a substantive approach of thermodynamics production in economy nor an ecological economics. Hence the discussions in \cite{Roegen, rosser, jayne} (where entropy is considered substantive and  the discussions are on macroeconomics) are not the case for our work.  A brief review article on this topic is, for example, \cite{Glucina}.
Further discussion on comparative aspects of physics and economics can be seen in  \cite{BGDemand, BGThesis}.

%%%%%%%%%%%%%%%% begin figure %%%%%%%%%%%%%%%%%%%
\section{Truly endogenous functions and EoS effect structures}
\begin{figure}[t]  \begin{center}
\includegraphics[width=13.7cm,height=3.1
cm,angle=0]{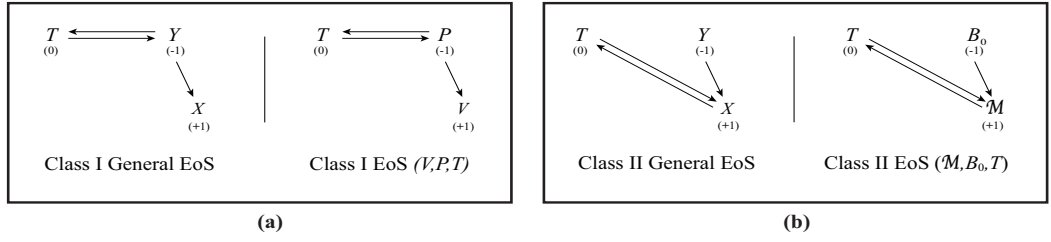}  \end{center}
\caption{Effect Structure Diagrams: (a) Class I and (b) Class II in general form and in specific form, i.e. hydrostatics and paramagnetics. Total degrees at each node, $k^{\rm tot}$ are shown as $(+1),(-1),(0)$.  \label{figClass1and2}}
 \end{figure}

Principles of equilibrium thermodynamics comprise of two major parts, the four postulated laws and the empirical EoS.
 Considering a simple hydrostatics system with EoS space of pressure (intensive), volume (extensive) and absolute temperature $\{V, P, T\}$, we notice that, among these three variables, (1) changing in $P$ is the cause of change in $T$ directly and changing in $P$ is the cause of change in $V$ directly, i.e. $P$ can affect $T$ directly and $P$ can affect $V$ directly.
(2) $T$ can affect $P$ directly. (3) $T$ can not affect $V$ directly, but only via the change in $P$ due to $T$ and (4) $V$ affects neither $T$ nor $P$ directly \cite{BG3}. This can be viewed and explained microscopically with mechanism of momentum exchanging at molecular level.
External factors (apart of variables of EoS space $\{V, P, T\}$), could influence the change in $P$ or $T$ (such as external force or external heat source) but $V$ can not be affected directly by external factors. $V$ can only be affected directly by changing in $P$ only\footnote{Note that here we do not restrict the possibility that extensive coordinate $X$ could not take external effect as it was restricted in \cite{BG3, BGDemand, BGThesis} of which has to be revised.}. We shall  call external effects, which are not of the variables of the EoS space $\{X, Y, T\}$, {\it exogenous}. Functions representing each direct effect, which is internal among $\{X, Y, T\}$, shall be defined as  {\it truly endogenous functions}. As written in textbooks, for example, $V = V(P,T)$ is in fact $V = V \circ P(T) =  V(P(T))$. Hence $T$ and $V$ are not truly endogenously related, but they are related via $P$. Therefore, all truly endogenous functions (tilde sign) of the $\{V, P, T\}$ system are:
\be
V = \tilde{V}(P), \;\;  P = \tilde{P}(T), \;\;   T = \tilde{T}(P)\,.
\ee
Argument of each function is the truly endogenous cause and the value of function is the truly endogenous effect.
Arrows represent truly endogenous function as directed graphs pointing from cause to effect (Figure \ref{figClass1and2}). Diagrams with double arrows linking $T$ and $Y$ are said to be of Class I. In paramagnetic substance, space of EoS is $\{ \mathcal{M}, C, B_0\} $ (paraelectric EoS resembles the paramagnetic case)
where $\mathcal{M}$ is magnetization (extensive), $C$ is the Curie constant and $B_0$ is external magnetic field intensity (intensive).  In this space one can embed a Class II EoS, $g(\mathcal{M}, B_0, T)  =  0$   or  $ \mathcal{M}= C B_0/T$ (simple paramagnetic substance).
All truly endogenous functions of paramagnetics EoS are:
$
\mathcal{M} = \tilde{\mathcal{M}}(T), \;\;  T = \tilde{T}(\mathcal{M}), \;\;   \mathcal{M} = \tilde{\mathcal{M}}(B_0)\,.
$
The class II EoS has double arrows linking $T$ and $X$ as in Figure \ref{figClass1and2}.
%%%%%%%%%%%%%%%%%%%%%%%%%
%%%%%%%%%%%%%%%%%%%%%%%%%%
Many of known physical EoS fall into these two classes. We inductively form {\it effect structure rules} describing this character of the  EoS, $g(X_i,Y_i,T) = 0$. These rules can be deployed as an additional criterion or a standard in determining the EoS status of an empirical formula. These are \footnote{Here we give corrections to the rules proposed earlier in \cite{BG3, BGDemand, BGThesis}}
\begin{enumerate}
  \item EoS $g(X,Y,T) = 0$ (of $n=1$) has effect structure in form of directed graph with three nodes and three directed links (arrows).
  \item There must be at least one arrow pointing $Y \rightarrow X$.
  \item Total degree of the node $X$ is $k_{X}^{\text{tot}} = +1$.
\end{enumerate}
These rules should give hints to mathematical and physical analogies to  EoS of other types of system.

\section{Demand-side economy}
\subsection{Manifold space, Carathéodory’s axiom and the first law}
We revise and sharpen previous work in \cite{BGDemand, BGThesis} where mechanical pair are $Q^{\d}$ of single commodity type ($X$ coordinate) and negative value of price, $- Pr$ (as $Y$ coordinate)\footnote{The minus sign is in the same spirit with hydrostatics system where $Y$ is $-P$.}.
$Q^{\d}$ has unit of commodity amount. Typically, $Y$ has a nature of influence per unit entity such as unit of charge or unit of area. Therefore $Pr$ is the cost of commodity per unit amount of commodity, i.e.
$
Pr = ({\text{total\;cost\;of\;commodity}})/({\text{quantity\;of\;commodity}})\,.
$
For $n$ types of commodity, there are mechanical pairs $(Q^{\d}_i, Pr_i)$ where $i = 1 \ldots n$. This is  $2(n+1)$ tuples manifold space $\{X_i, Y_i, S, T\}$ of which degrees of freedom are left with $(2n+1)$ at equilibrium as $S$ is fixed, hence $\mathcal{M} = \{X_i, Y_i, T\}$\footnote{Example of a multiple EoS is such as a fluid paramagnetics substance with $X_1 \equiv V$, $X_2 \equiv \mathcal{M}$, $Y_1 \equiv -P$, $Y_2 \equiv B_0$ }.
In microeconomics, the manifold space is consumption quantities of $n$ types of commodity $\{ Q^{\d}_1, Q^{\d}_2, \ldots, Q^{\d}_i \ldots Q^{\d}_n \}$. Utility, $\mathcal{U}$ functions are presented with set of infinite indifference curves (I.C.).  $Q^{\d}_i$  are coordinate variables, hence utility is considered in microeconomics as a state function, $\mathcal{U} = \mathcal{U}(Q^{\d}_i)$ for at least two  type of commodity, i.e. $n \geq 2$.
Constrained optimization with known budget equation at one level of $\mathcal{U}$ can settle an equilibrium price. Spirit of thermodynamics is quite different.  In thermodynamics, optima are achieved with the first law and the second law. Thermodynamics EoS should provide the value of $Pr_i$ and $Q^{\d}_i$ to be definite at one level of consumers' economics temperature, $\varphi$, to be defined later. We argue that utility should be acquired as heat transfer term rather than as a state function. In thermodynamics formulation, despite of having single type of commodity, equilibrium price and utility can be obtained.

Existence of infinite thermal states is postulated with Carathéodory’s axiom. The axiom is applied to system of consumers regarding to four
common characteristics of thermodynamics and microeconomics as stated in section
\ref{Intro}.
Carathéodory’s axiom strictly postulates temperature from infinite thermal states of mechanical pairs,
$
f(Pr^1_i, Q^{\rm d, 1}_i)  = \varphi^1, \;\; f(Pr_i^2, Q^{\rm d,2}_i)  = \varphi^2 \;\ldots\,,
$
as well as a set of infinite entropic  states to exist,
$
s(Pr^1_i, Q^{\rm d, 1}_i)  = S^{\d, 1}, \;\; s(Pr^2_i, Q^{\rm d,2}_i)  = S^{\d, 2} \;\ldots\,.
$
Superscripts denote states of variables. This allows EoS $g(Q^{\d}_i, Pr_i, \varphi) = 0$ to exist.
In both disciplines, knowledge of $(d-1)$-dimension surface embedded in $d$-dimension space is a key idea of {\it integrability problem}.
Exact form of the surface implies fundamental constraint equations.
In microeconomics, as in \cite{sousa2006} which reports $\pi_i = \pi_i(X_1,\ldots X_i \ldots X_n)$,
the surface,  is $(2n-1)$-dimensional embedded in $2n$-dimensional space,
 $\{Pr_i, Q^{\d}_i\}$. The surface, $Pr_i =  Pr_i(Q^{\d}_1, \ldots, Q^{\d}_i, \ldots, Q^{\d}_n)$ is indeed,
$
f(Pr_i, Q^{\d}_i) = 0\,,
$
 where there must be at least two commodities, i.e. $n \geq 2$. This implies budget constraint equation as
 $ I = \sum_{i=1}^{n \geq 2} Pr_i Q^{\d}_i, $
 which is the fundamental constraint equation.
 On the other hand, in thermodynamics formulation, the surface is $2n$-dimensional embedded in $(2n+1)$-dimension space. In thermodynamics formulation of demand-side economy, the surface is the EoS,
$
g(Q^{\d}_i, Pr_i, \varphi) =  0\,.
$
In thermodynamics formulation, even one type of commodity is allowed, i.e. $n \geq 1$. According to Carathéodory’s axiom, this implies the first law,
$
\d \mathcal{W} \;=\; \sum_{i=1}^{n} (-Pr_i) \, \d Q^{\d}_i \:+\: \varphi\, \d S^{\d} \:+\: \mu^{\d}_j \,\d N^{\d}_j   % \label{firstlaw}
$
which is the fundamental conservative constraint. The consumers' temperature $\varphi$, wealth function $\mathcal{W}$ and consumers' entropy, $S^{\d}$ are to be defined later.  When number of consumers is changed, thermodynamics formulation can also account for it. Energy transfer from matter exchange, $\mu_j\, \d N_j$ (where $\mu_j$ is chemical potential ($j = 1, \ldots, m$) and $N_j$ is particle number of a type $j$ of substance) can also be applied. Number of $j$th-group consumers can be analogously related to number of particle of type $j$ substance, $N_j$. This is the same spirit with consumers' chemical potential, $\mu^{\d}_j$ for distinct group of consumers. It makes sense to say that a physical substance is homogeneous, but, in social systems,  each individual agent is different from others.  Although, system of distinct consumers is not homogenous, however it is reducible as system of gas molecules. In a closed homogeneous physical system, $j=1$ and $\d N = 0$. Behaviors of whole closed consumers' economy  are considered averagely and aggregately, hence closed system of consumers is treated in the same spirit as a closed homogeneous physical system, $j=1$ and $\d N^{\d} = 0$. Therefore the term, $\mu^{\d}_j\, \d N^{\d}_j$, is not needed for the first law.

 \subsection{Value, wealth, temperature, generalized utility, expenditure and the Euler's equation}
 In our concept, {\it value} expressed in money unit is analogous to energy. Value is in form of assets (e.g. cash, stock, car, house, etc.) and in forms of utility, i.e. satisfaction, opportunity or benefit of usage, etc.
Wealth function, $\mathcal{W}$ at one instance  includes all types of value possessed by a whole system of consumers subtracted by total cost of expenditure,
\be
\mathcal{W} \,=\, -\big({\text{total\;expenditure}}\big) + \big[{\text{(total\;ownership\;of\;assets)\:+\:(total\;utility)}}\big].   \label{wealthpart}
\ee
Wealth plays the role of conservative quantity in the first law just as $U$.  As  $X_i$ is $Q^{\d}_i$, $Y_i$ is $-Pr_i$ and $T$ is $\varphi$, the thermodynamics space is hence $\{ Q^{\d}_i, Pr_i, \varphi \}$ therefore $ \mathcal{W}$ is a field potential,
$
\mathcal{W} = \mathcal{W}(Q^{\d}_i, Pr_i, \varphi)\,.
$
The first law (\ref{firstlaw}) suggests natural coordinates $\{Q^{\d}_i, S^{\d}, N^{\d}\}$  for $\mathcal{W}$, hence
$
\mathcal{W} = \mathcal{W}(Q^{\d}_i, S^{\d}, N^{\d})\,.
$
The $2n$-dim EoS surface is a constraint reducing one degree of freedom of $\mathcal{W}$ to $2n$.
Analogy to absolute temperature is {\it average wealth per each consumer}, $\varphi$. As absolute temperature of idealized systems is $T \propto U/N$, i.e. concentration of internal energy per particle, hence
$
\varphi = {\mathcal{W}}/{N^{\d}}\,,
$
where $N, N^{\d}$ are number of molecules or consumers respectively.
Energy transfer in form of work, $\delta W = \sum^n_{i = 1} Y_i \d X_i$  is
analogous to the value transfer in form of {\it expenditure}:
$
\delta W^{\d}_i = \sum^n_{i = 1}  (-Pr_i)\, \d Q^{\d}_i.
$
Demand-side economy has two activities: spending and consuming. Reversing consumption process does not give any production and spending (buying) is not a reverse of selling. In a piston of gas, expansion and contraction are reversed to each other since both are the effect of microscopic momentum exchange. Consumption and  production do not resemble this. Although, the consumers might be the same people who work for production but both activities are solely different. Consumers only spend and consume hence expenditure  (work done by the system) always comes with minus sign in front of $Pr$. Note that if $\d Q^{\d}_i < 0$, i.e reducing of consumed quantity, $\delta W^{\d}_i$ can be positive.
 The other value transfer is the {\it generalized utility}, $\delta \mathcal{Q}_{\text{util}}^{\d}= \varphi \d S^{\d}$ which resembles heat, $\delta \mathcal{Q} = T \d S$ for a reversible process. Therefore the first law  reads
 $
\d \mathcal{W}  \;=\;  \delta W^{\d} + \delta \mathcal{Q}_{\rm util}^{\d}\,. $
Generalized utility includes many aspects of value, for examples\footnote{When spending, consumers gain satisfaction (or utility).  Satisfaction should be included in the conservation law (first law).  Considering only asset ownership in the conservation law does not fulfill the picture.},
 \begin{enumerate}
  \item ownership status of commodities which becomes the consumers' assets in physical forms or non-physical forms (e.g. money, land, car, stock, business, firm, etc.)\footnote{Assets are not considered as values but it is ownerships status (of the assets) that are the values.}.
  \item benefit gained from the rights of using or consuming
  \item satisfaction, pleasures or contentment
  \item  opportunity to invest.
\end{enumerate}
Generalized utility can be transferred in both ways, i.e. receiving and giving. Being given an ownership of a house by someone for free is considered as heat transfer without work done, i.e. without any expenditure. Our generalized utility function is inexact differential and is not a state function while microeconomics' utility function, $\mathcal{U}$ is viewed as a state function in microeconomics\footnote{It has been
understood that utility is a state function and is a stock variable, however utility happens only temporarily at the moment of consumption, therefore it  should be flow variable.}.  Generalized utility is hence a value transfer just as heat transfer and is notified with $\delta$ for its inexact differential nature. In thermodynamics, considering simple case of  $m=1$, i.e. single type of substance, the
Euler's equation is  $U = \sum_{i=1}^{n} X_i Y_i + TS + \mu N$  and the Gibbs free energy is  $G = \mu N$. Recall $G = U - \sum_{i=1}^{n} X_i Y_i - TS$ and this is $G = U + PV - TS$ in a hydrostatics system.  In consumers system, the Euler's equation is
$
\mathcal{W} \:=\: \sum_{i=1}^{n} (-Pr_i)\,Q^{\d}_i \,+\, \varphi\,  S^{\d} \,+\, \mu^{\d} N^{\d}
$
Compared to equation (\ref{wealthpart}), $\sum_{i=1}^{n} (-Pr_i)\,Q^{\d}_i  $  is the  $-(\text{total\: expenditure})$,
$\varphi\,S^{\d}$ is the total generalized utility which is sum in the square brackets. Keep in mind that generalized utility, $\delta \mathcal{Q}_{\text{util}}^{\d}$ includes four aspects as mentioned above. The consumers' Gibbs function,
$
G^{\d}  \: = \:\mu^{\d} N^{\d} \:=\: \mathcal{W}   + \sum_{i=1}^{n} (Pr_i)\,Q^{\d}_i  -  \varphi\,  S^{\d}
$
is the total wealth added by spent total expenditure and subtracted by total generalized utility. Hence $G^{\d}$ is the original wealth if the system
does not spend and does not have any generalized utility in any forms. In other words, Gibbs function is total value that consumers originally occupy as they initially come into existence or into our consideration. Consumers' chemical potential is hence the average concentration of $G^{\d}$ per consumer, $\mu^{\d} = G^{\d}/N^{\d}$.
Consumers' enthalpy ($\mathcal{H}^{\d}$), Helmholtz function ($\mathcal{F}^{\d}$) and grand potential ($\Phi^{\d}$) can also be obtained under Legendre transformation. In natural coordinate arguments, these are $\mathcal{W} = \mathcal{W}(Q^{\d}, S^{\d}, N^{\d})$ and $G^{\d} = G^{\d}(Pr, \varphi, N^{\d})$,
  $\mathcal{H}^{\d} = \mathcal{H}^{\d}(Pr, S^{\d}, N^{\d})$,   $\mathcal{F}^{\d}  =  \mathcal{F}^{\d}(Q^{\d}, \varphi, N^{\d})$,  $\Phi^{\d} =  \Phi^{\d}(Q^{\d}, \varphi, \mu^{\d}).$ In the same spirit with thermodynamics, enthalpy is $\mathcal{H}^{\d} = \mathcal{W}+ \sum_{i=1}^{n} Pr_i\,Q^{\d}_i$, i.e. sum of total wealth and positive value of total expenditure. Helmholtz function is $\mathcal{F}^{\d} = \mathcal{W}  -  \varphi S^{\d}$, i.e.  total wealth subtracted by value of total generalized utility. Grand potential, $\Phi^{\d} = \mathcal{F}^{\d} - \mu^{\d} N^{\d} = - \sum_{i=1}^{n} Pr_i\,Q^{\d}_i$, i.e. cost of total expenditure.

\subsection{Consumers' entropy, the second law, the zeroth law, value equilibrium, value contact and generalized consumer surplus (deficit)}
The variable $S^{\d}$ is the consumers' entropic function. The change in consumers' entropy is $\d S^{\d} =  {\delta \mathcal{Q}^{\d}_{\rm util}}/{\varphi}\,. $ Generalized utility, $\delta \mathcal{Q}_{\text{util}}^{\d}$ (heat term) is not the quantity to be maximized. In our concept, inequality of the second law maximizes $S^{\d}$. This is possible even with only one type of commodity and $\delta \mathcal{Q}_{\text{util}}^{\d}$ can be produced even without any spent expenditure $\delta W^{\d} = 0$.  It only follows $\Delta S^{\d} \geq 0.$ Level of consumers' entropy $S^{\d}$ is interpreted as generalized utility per unit of average personal wealth.   For the same commodity, the richer (higher average personal wealth) have less consumers'entropy in consumption or using a product or service. Simply saying, the richer are less excited or less happy when using the same product or service compared to the poorer. Because it is easier for the richer to acquire the product or the service.

The zeroth law helps defining {\it value equilibrium} (thermal equilibrium) between two groups of consumers.  The value equilibrium is the condition when average personal wealth, $\varphi$ of the two groups are equal and steady. It may be hard to state precisely about equality of $\varphi$ of two groups in real-world society of consumers with high diversity of their economics status.  However, as discussed earlier, the system is considered as a whole averagely and aggregately. Supposing that the two systems are smallest, i.e. there are only two consumers with different $\varphi$.  Value (thermal) contact is the situation when they can share their values  
via marriage, partnership, adoption, becoming family or united as one household. Value equilibrium, $\varphi_1=\varphi_2$, is achieved slowly.
 Generalised utility (heat), in money unit, is transferred to each other in the value contact.  Assets ownership (an aspect of generalized utility)  is transferred without expenditure (no work term).

\begin{figure}[t]  \begin{center}
\includegraphics[width=8.5cm,height=3.7
cm,angle=0]{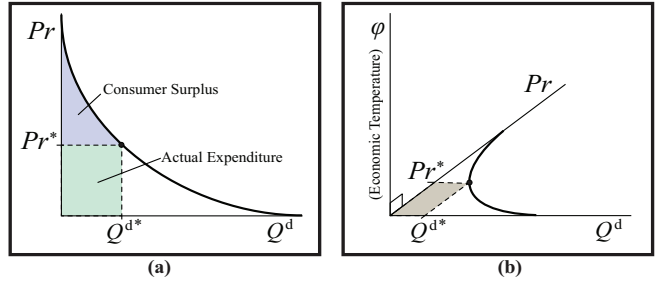}  \end{center}
\caption{(a) Schematic demand curve: showing consumer surplus in microeconomics (b) Space of thermodynamics formulation of economics suggests that consumer surplus should be evaluated off the $ Pr - Q^{\d} $  plane to the $\varphi$ axis. \label{figsurplus}}
 \end{figure}

  In microeconomics, consumer surplus is an exceeding utility at equilibrium price obtained by consumers. This is presented as area under demand curve and above the equilibrium price in Figure \ref{figsurplus}. In thermodynamics picture, utility is of heat nature and it does not make sense to consider utility as area under the curve in mechanical pair $\{Q^{\d}, Pr\}$ plane.
  In our  concept, since consumer surplus is an exceeding total generalized utility obtained more (or less) than the cost of expenditure spent. Hence generalized consumer surplus (or deficit) is defined as
$ \Psi  \equiv   \varphi^*  S^{\d *}     - Pr^*\, Q^{\d *}\,,  $
where $*$ denotes equilibrium state. This is in fact a difference between total wealth function and Gibbs function, $\Psi = \mathcal{W} - G^{\d}\,$. Since Gibbs function is the initial wealth before any process, i.e. before spending or acquiring (losing) of generalized utility, hence $\Psi$  is  the gained or lost value of wealth from the initial wealth after the process completes.

\subsection{Processes and the third law}
For simplicity, instead of discussing processes in general $\{X_i,Y_i\}$ coordinates, we make our analogy to processes in hydrostatics system. {\it Adiabatic process} (zero heat, $\delta \mathcal{Q}_{\text{util}}^{\d} = 0$) gives
$ \d \mathcal{W} = \delta W^{\d}\,,$ which implies spending without gaining any generalized utility (asset ownership or other utility). This is as idealized as in thermodynamics. 
Economics version of {\it isovolumic process} (zero work or zero spending), 
$ \d \mathcal{W} = \delta \mathcal{Q}^{\d}_{\rm util}\,.$ This is a transferring of generalized utility without any spending. This case is hence the value (thermal) contact.  In case of consumers with constant average personal wealth, {\it isothermal process} (constant $\varphi$) is in fact
Adam Smith’s price mechanism. Therefore the law of demand in microeconomics is a specific case of our formulation.
{\it Isobaric process} is analogous to constant price process, to be named, {\it isoprice}. Demand quantity depends on only average personal wealth. If consumers are  averagely richer from exogenous effects, demand should increase. On the other hand, if they are poorer from exogenous effects, demand should decrease.

In analogous with the third law of thermodynamics, it takes infinite economic isothermal and economic adiabatic steps to reach zero average personal wealth, $\varphi = 0$. At this state, generalized utility per personal wealth, $S^{\d}$ is postulated to be zero.  Wealth includes value such as happiness, hence there are infinite steps to remove it away completely. If such state with $\varphi = 0$ exists, i.e. no value nor happiness in any forms, nor asset ownership, nor even passion of happiness, the state of zero $S^{\d}$ (generalized utility per personal wealth) is postulated to exist. Economics agents are in existence not because of the existence of only assets. For them, to spend or to share the value of assets, they need to have memory of happiness or having passion. Economics agents realize that they live for something valuable. The $\varphi = 0$ state of a consumer is when consumer gives up any types of value in life, resembling the nirvana state in Buddhism philosophy.

\begin{figure}[t]  \begin{center}
\includegraphics[width=7.1cm,height=2.9
cm,angle=0]{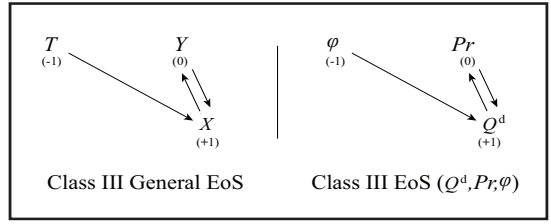}  \end{center}
\caption{Effect structure diagrams for Class III EoS  in general form and in system of consumers. Total degrees at each node, $k^{\rm tot}$ are shown as $(+1),(-1),(0)$.  \label{figClass3}}
 \end{figure}

\subsection{Class III diagram and adjacency matrices of the EoS graphs}
With knowledge in conventional microeconomics together with concept of effect structure rules proposed above, the effect structure diagram for consumers' economy is presented in Figure \ref{figClass3}.  The diagram is a new class with two opposite-direction arrows linking $Y$ and $X$. We regard it as Class III diagram\footnote{We revise our effect structure rules reported earlier in \cite{BGDemand}. Note that in \cite{BGDemand} some subclass diagrams of Class III are allowed. Here in this present work, the rules are revised and subclass diagrams are no longer allowed.}.
Effect $\varphi \rightarrow Q^{\d}$ is the isoprice (isobar) process. The effects $Pr \rightleftarrows   Q^{\d}$ are the microeconomics law of demand (isothermal process).

EoS effect structure diagrams can be represented more formally with adjacency matrices. Let $A_{ab}$ be network adjacency matrix element where $a, b$ are number of row and column.  The element $A_{ab} $ is $1$ if there is a directed link from node $a$ to node $b$. Given $a,b=1$ for $X$ node; $a,b=2$ for $Y$ node and $a,b=3$ for $T$ node in the diagrams in Figures \ref{figClass1and2} and \ref{figClass3},
adjacency matrix for Class I, II and III are,
\be
 A_{ab, \text{I}} \; \; = \;\;
\l[ \begin{array}{cccc}
 0 & 1           & \; 0 \;   \\
0 &  0           & \; 1 \;   \\
0 & 1           & \; 0 \;   \\
\end{array} \r]\,,    \;\;\;\;\;\;\;\;\;\;
 A_{ab, \text{II}}  \; \; = \;\;
\l[ \begin{array}{cccc}
 0 & 1           & \; 1 \;   \\
0 &  0           & \; 0 \;   \\
1 & 0           & \; 0 \;   \\
\end{array} \r]    \;\;\;\text{and} \;\;\;
 A_{ab, \text{III}} \; \; = \;\;
\l[ \begin{array}{cccc}
 0 & 1           & \; 1 \;   \\
1 &  0           & \; 0 \;   \\
0 & 0           & \; 0 \;   \\
\end{array} \r]   \label{ClassAd}
\ee
This graph analysis is of interest for further investigation.

%%%%%%%%%%%%%%%%%%%%%%%%%

\section{Econometric modelling  of consumers' EoS}
In general form, consumers' EoS is  $g(Q^{\rm d}, Pr, \varphi) = 0$. To evaluate the EoS, we consider total differential of demand function,
$
\d Q^{\rm d}   =    \l({\p Q^{\rm d} }/{\p \varphi}   \r)_{Pr} \d \varphi   +   \l({\p Q^{\rm d} }/{\p Pr}   \r)_{\varphi} \d Pr\,.
$
We define
$
\beta_{Pr}   \equiv    \l({1}/{Q^{\rm d}_0}\r)  \l({\p Q^{\rm d}}/{\p \varphi}  \r)_{Pr}   =   {E^{\rm d}_{\varphi}}/{\varphi_0}\,$ 
and 
$
\kappa_{\varphi}   \equiv    -\l( {1}/{Q^{\rm d}_0}\r)   \l({\p Q^{\rm d}}/{\p Pr}  \r)_{\varphi}  =   -{E^{\rm d}_{Pr}}/{Pr_0},
$
where $E^{\rm d}_{\varphi}$ and  $E^{\rm d}_{Pr}$ are elasticity of demand to personal wealth and elasticity of demand to price (as in microeconomics). Elasticities may not be constant and could depend on other factors as of coefficients, $\beta, \kappa$ in hydrostatics EoS.  Straightforwardly, from the total differential, 
$
Q^{\rm d}(Pr, \varphi)  =   Q^{\rm d}_0 \l[ 1 + \beta_{Pr}(\varphi - \varphi_0)  -  \kappa_{\varphi}(Pr - Pr_0)  \r] \,,  
$
which is the EoS of the consumers' system. For econometric modelling, we define $
Y  \equiv  {Q^{\rm d}  }/{Q^{\rm d}_0}\,, 
X_1  \equiv  \varphi - \varphi_0\,,   
X_2  \equiv  Pr - Pr_0\,,
$ and the EoS is hence,
$  Y \;=\;    1 + \beta_{Pr}X_1  -  \kappa_{\varphi} X_2  +   u \,, $
where $u$ is an error term.  One can perform econometrics regression analysis of this EoS equation against data such that coefficients $\beta_{Pr}$ and $\kappa_{\varphi}$ are determined.

\section{Conclusions}
We discuss characters of social and physical systems. Four common natures of thermodynamics and economics are collective and aggregated nature, equilibrium determined with dual states, conservative constraints and optimization preferences.  Economics systems are speculated to have further thermodynamics aspects such as existence of thermal and entropic states, EoS and so forth.  We discover effect structure of EoS in general and propose diagrams and rules for justifying empirical laws to EoS status. This  is applied  to demand-side economy and we are able to find EoS of consumers system.  EoS are classified into three classes represented with effect structure diagram and adjacency matrices.  We identify space of thermodynamics formulation of demand-side economics and compare it to microeconomics space. We found that {\it value} is fundamental entity in analogy with energy. The other analogies are total wealth as internal energy, expenditure as work done, generalized utility (newly defined) as heat, average personal wealth as temperature. We interpret analogous thermodynamics potentials such as Gibbs function and so forth. We criticize the concept of consumer surplus in microeconomics and propose new concept of generalized consumer surplus 
(deficit).  Our theory includes law of demand as a specific case, i.e.  isothermal process.  Thermal (value) contact, consumers's analogous entropy and all four laws of thermodynamics are interpreted in economy context. Absolute zero temperature (with zero entropy) state in the third law corresponds to the state when consumers give up all values in life, resembling the nirvana state in Buddhism philosophy. At last, we propose econometric method for modelling consumers' EoS with data.

\section*{Acknowledgments}

This research project is supported by Mahidol University (grant no. MRC-MGR 04/2565).


\begin{thebibliography}{9}



% -- Articles in journals    -- %

\bibitem{BGDemand} B. Gumjudpai, Thermodynamics formulation of economics (an extended abstract), in: \emph{International Conference on Thermodynamics 2.0: where natural sciences meet social sciences}, Worcester, Massachusetts, 22-24 June, 2020,
    https://arxiv.org/pdf/2012.01505v2.
\bibitem{BGThesis}
B. Gumjudpai, \emph{Thermodynamics formulation of economics}, M.Econ. (Financial Economics) thesis, National Institute of Development Administration (NIDA), Bangkok, 2019.

  \bibitem{BG3}  B. Gumjudpai and Y. Sethapramote, Effect structure in physics and hints to economics equation of state,  \emph{J. Phys. Conf.  Ser.}  \textbf{1380} (2019), 012167.

 \bibitem{BG2}  B. Gumjudpai and Y. Sethapramote, Nature of thermodynamics equation of state towards economics equation of state, in:
   \emph{The 11th Silpakorn University Research Fair}, Nakhon Pathom, Thailand, 13-14 June 2019, https://arxiv.org/pdf/1907.07108v1.

 \bibitem{BG1}  B. Gumjudpai, Towards equation of state for a market: a thermodynamical paradigm of economics,  \emph{J. Phys. Conf.  Ser.}  \textbf{1144} (2018), 012181, https://arxiv.org/pdf/1807.09595v2.



\bibitem{munster} A. M\"{u}nster,  \emph{Classical Thermodynamics}, Wiley-Interscience, London, 1970.


\bibitem{cara} A. Land\'{e}, Die Carath\'{e}odory'sche axiomatik, in \emph{Geiger-Scheel Handbuch der Physik  vol. 9}, Springer-Verlag, Berlin (1926) chapter 4.
 \bibitem{cara2}
H. A. Buchdahl, Simplification of a proof of Carathéodory’s theorem, \emph{Am. J. Phys.}  \textbf{23}  (1955), 65-66.
 \bibitem{cara3}
L. A. Turner, Simplification of Carathéodory’s treatment of thermodynamics, \emph{Am. J. Phys.} \textbf{28} (1960),
781-786.
 \bibitem{cara4}
 H. A. Buchdahl, \emph{The Concepts of Classical Thermodynamics}, Cambridge University Press, London, 1966.

\bibitem{debrue} G. Debrue, \emph{Theory of Value}, Yale University Press, New Haven and London, 1972.
\bibitem{Mas} A. Mas-Colell, \emph{The Theory of General Economic Equilibrium: A Differentiable Approach}, Cambridge University Press, London, 1985.

\bibitem{Smit} E. Smith and D. K. Foley, Classical thermodynamics and economic general equilibrium theory, \emph{J. Econ. Dyn. Contr.} \textbf{32} (2008), 7.
\bibitem{Fish} I. Fisher, Mathematical investigations in the theory of value and prices, doctoral thesis, Yale University,
in \emph{Trans. of the Connecticut Acad. vol. 9}, New Haven (1892).



\bibitem{walras}  L. Walras, Economique et m\'{e}canique, \emph{Bull. Soc. Vaudoise Sci. Nat.} \textbf{45} (1909) 313.
\bibitem{Lisman}  J. H. C. Lisman, \emph{Econometrics, Statistics and Thermodynamics}, The Netherlands Postal and Telecommunications Services, Holland,  1949.

 \bibitem{Sasl} W. M. Saslow,  An economic analogy to thermodynamics,  \emph{Am. J. Phys.}  \textbf{67} (1999), 1239.

  \bibitem{Samu} P. A. Samuelson,  Structure of a minimum equilibrium system, in:  \emph{Essays in Econ. Econometrics: A Volume in Honor of Harold Hotelling}, University of North Carolina Press, (1960).     %R. W. Pfouts (Ed.)

 \bibitem{sousa2006}
T. Sousa, and  T. Domingos,  Equilibrium econophysics: a unified formalism for neoclassical economics and equilibrium thermodynamics,  \emph{ Physica A}  \textbf{371} (2) (2006),  492-512.

\bibitem{Colon}
L. B. Arroyo Col\'{o}n, \emph{A thermal model of the economy}, masters' thesis, University of Puerto Rico at Mayagüez, 2010.

\bibitem{Roegen}
N. Georgescu-Roegen,   \emph{The Entropy Law and the Economic Process}, Harvard University Press, Cambridge, MA, 1971.

\bibitem{rosser} J. B. Rosser Jr.,  Entropy and econophysics, \emph{Eur. Phys. J. Special Topics}  \textbf{225} (2016), 3091-3104.

\bibitem{jayne}
E. T. Jaynes, How should we use entropy in economics?, preprint (1991), https://bayes.wustl.edu/etj/articles/entropy.in.economics.pdf.


\bibitem{Glucina}
M. D. Glucina and K. Mayumi, Connecting thermodynamics and economics,
\emph{Ann. N.Y. Acad. Sci.} \textbf{1185} (2010), 11-29.





\end{thebibliography}
\end{document}